\documentclass[seceq]{ptptex}

\newcommand{\be}{\begin{equation}}
\newcommand{\ee}{\end{equation}}
\newcommand{\bea}{\begin{eqnarray}}
\newcommand{\eea}{\end{eqnarray}}

\newcommand{\eq}[1]{Eq.~(\ref{eq:#1})}
\newcommand{\sect}[1]{Sec.~\ref{sec:#1}}
\newcommand{\appen}[1]{Appendix~\ref{sec:#1}}

\newcommand{\bra}{\langle}
\newcommand{\ket}{\rangle}
\newcommand{\del}{\partial}
\newcommand{\vecx}{\vec{x}}

\newcommand{\srmg}{\sqrt{-g}}
\newcommand{\srmgz}{\sqrt{-g^{(0)}}}

\newcommand{\nw}{\mathfrak{w}}

\newcommand{\cO}{{\mathcal{O}}}

\title{The shear viscosity of holographic superfluids 
}

\author{
Makoto Natsuume
\footnote{E-mail: makoto.natsuume@kek.jp }
and Masahiro Ohta
\footnote{E-mail: ohtam@post.kek.jp }
}

\inst{KEK Theory Center, Institute of Particle and Nuclear Studies, \\
High Energy Accelerator Research Organization (KEK), \\
1-1 Oho, Tsukuba, Ibaraki, 305-0801, Japan \\ 
and \\
Department of Particle and Nuclear Physics, \\
The Graduate University for Advanced Studies (SOKENDAI), \\
1-1 Oho, Tsukuba, Ibaraki, 305-0801, Japan}

\abst{
We study the ratio of the shear viscosity to the entropy density for various holographic superfluids. For the $s$-wave case, the ratio has the universal value $1/(4\pi)$ as in various holographic models. For the $p$-wave case, there are two shear viscosity coefficients because of the anisotropic boundary spacetime, and one coefficient has the universal value. For the $(p+ip)$-wave case, the existing technique is not applicable since there is no tensor mode of metric perturbations which decouples from Yang-Mills perturbations. Our results indicate that the shear viscosity does not show a singular behavior at the critical point for holographic superfluids. 
}

\PTPindex{121}  

\begin{document}

\maketitle

\section{Introduction}\label{sec:intro}

In the studies of holographic superconductors/superfluids \cite{Gubser:2008px,Hartnoll:2008vx,Gubser:2008zu,Gubser:2008wv,Hartnoll:2008kx,Herzog:2008he,Basu:2008st} (See, {\it e.g.}, Refs.~\citen{Hartnoll:2009sz,Herzog:2009xv,Horowitz:2010gk} for reviews), one often uses numerical computations or some approximations. This is because the holographic superfluids are Einstein-matter systems and it is in general difficult to solve such systems. One approximation often employed is the ``probe approximation," where the backreaction of matter fields onto the geometry can be ignored. While the approximation is enough to see the phase transition and to compute properties such as the conductivity, gravitational properties of these systems, in particular analytic results are largely intact. It is desirable to obtain gravitational properties of these systems analytically. We investigate this issue in this paper. We study $\eta/s$, the ratio of the shear viscosity to the entropy density for holographic superfluids.

Studying $\eta/s$ is interesting from another point of view. According to the AdS/CFT duality, the ratio is universal, {\it i.e.},
\be
\frac{\eta}{s} = \frac{1}{4\pi}
\label{eq:univ}
\ee
in the large-$N$ limit. This holds for all known examples which have been studied (See, {\it e.g.}, Ref.~\citen{Natsuume:2007qq} and references therein). Even though there exists several arguments to support the universality \cite{Buchel:2003tz,Kovtun:2004de,Benincasa:2006fu,Iqbal:2008by}, it is still unclear why the universality holds microscopically and how generic the universality is. The holographic superfluids provide yet another example of the universality. 

The holographic superfluids exhibit a second-order phase transition. At high temperatures (normal phase), the bulk geometry is given by the standard Reissner-Nordstr\"{o}m-AdS black hole. The shear viscosity for the Reissner-Nordstr\"{o}m-AdS black hole has been computed in Refs.~\citen{Mas:2006dy,Son:2006em,Saremi:2006ep,Maeda:2006by}, and the universality has been shown already. Thus, we focus on the low temperature phase (superfluid phase).

Technically, the universality largely depends on the following two properties of the bulk theory:
\begin{enumerate}
\item One can use the Kubo formula to compute $\eta$ and carry out the tensor mode computation of gravitational perturbations. There are no other fields which transform as a tensor even if matter fields are present.

\item The entropy density is given by the Bekenstein-Hawking formula as long as the gravitational action takes the Einstein-Hilbert form.
\end{enumerate} 

In this paper, we consider three class of holographic superfluids, the $s$-wave, $p$-wave, and $(p+ip)$-wave holographic superfluids in $(d+1)$-dimensional bulk spacetime. Our results are summarized as follows
\footnote{
See Ref.~\citen{Buchel:2010wf} for the shear viscosity of another holographic superfluid.}
:
\begin{enumerate}
\item[(i)] The $s$-wave holographic superfluids are described by Einstein-Maxwell-complex scalar systems \cite{Gubser:2008px,Hartnoll:2008vx,Hartnoll:2008kx}. In this case, the universality holds with a modification of the technique in Ref.~\citen{Benincasa:2006fu}.

\item[(ii)] The $p$-wave holographic superfluids are described by Einstein-Yang-Mills systems \cite{Gubser:2008zu}. In this case, the Yang-Mills field breaks the $SO(d-1)$ rotational invariance on the boundary theory, which has two implications. First, the hydrodynamic limit is not described by a single shear viscosity.%
\footnote{
In the context of the AdS/CFT duality, anisotropic shear viscosities have been computed for the noncommutative ${\cal N}=4$ plasma \cite{Landsteiner:2007bd}. }
Second, for $d=3$, one does not have a tensor mode which decouples from the Yang-Mills field. (Namely, item~1 of the above list fails.) As a result, the existing technique is not applicable. However, for $d \geq 4$, one has the $SO(d-2)$ invariance. In this case, a tensor mode exists, and the universality holds for the shear viscosity associated with the tensor mode.

\item[(iii)] The $(p+ip)$-wave holographic superfluid is described by the same system as the $p$-wave holographic superfluid (with $d=3$), but the symmetry breaking pattern is different \cite{Gubser:2008wv}. Although the metric keeps the $SO(2)$ invariance, the Yang-Mills field breaks the $SO(2)$ invariance. As a result, there does not exist the tensor mode which decouples from Yang-Mills perturbations and the existing technique is not applicable.

\end{enumerate}
Our results indicate that the shear viscosity has no singular behavior across the phase transition for holographic superfluids (See \sect{DCP}).

The plan of this paper is as follows. In \sect{s-wave}, we consider $\eta/s$ for the $s$-wave holographic superfluids.  In \sect{p-wave}, we consider anisotropic holographic superfluids, the $p$-wave and $(p+ip)$-wave holographic superfluids. For the $(p+ip)$-wave case, we identify the Yang-Mills perturbations which couple to the would-be tensor mode of metric perturbations. In \sect{discussion}, we discuss the implications of our results. We discuss the shear viscosity itself of holographic superfluids. For a $p$-wave superfluid, the entropy density $s$ has been obtained, so we can discuss one of shear viscosity coefficients itself. The shear viscosity coefficient in the superfluid state is lower than the one in the (unstable) normal state.

\section{The $s$-wave superfluids}\label{sec:s-wave}

\subsection{Background}

The $s$-wave holographic superfluids are described by Einstein-Maxwell-complex scalar system:
\be
\begin{split}
S_{s} = \frac{1}{16\pi G_{d+1}}\int d^{d+1}x \sqrt{-g} \bigg\{ 
       R - 2 \Lambda &- \frac{1}{4}K_{1}\left(|\Psi|^{2}\right) F^{MN} F_{MN} \\
       &- K_{2}\left(|\Psi|^{2}\right) \left\vert \nabla_M \Psi - iq A_M \Psi \right\vert^2 
       - V\left( | \Psi |^2 \right) \bigg\}~
\end{split}
\label{eq:action_s-wave}
\ee
with the ansatz
\bea
ds^2_{d+1} &=& - g_{tt}(r) dt^2 + g_{xx}(r) \sum_{i=1}^{d-1} dx_i^2 + g_{rr}(r) dr^2~,\label{eq:s-wave_metric} \\
A &=& A_t(r)dt~, \label{eq:s-wave_gauge}\\
\Psi &=& \Psi(r)~. \label{eq:s-wave_scalar}
\eea
Here, capital Latin indices $M, N, \ldots$ run through bulk spacetime coordinates $(t, x_i, r)$, where $(t,x_i)$ are the boundary coordinates and $r$ is the AdS radial coordinate. Greek indices $\mu, \nu, \ldots$ run though only the boundary coordinates. $K_{1}$, $K_{2}$ and $V$ are arbitrary real functions of the scalar field. This action includes not only the conventional $s$-wave holographic superfluids \cite{Gubser:2008px, Hartnoll:2008vx} but also generalized models \cite{Franco:2009yz, Franco:2009if, Herzog:2010vz}. We impose the regularity condition on the metric at the horizon $r=r_{h}$: 
\be
 g_{tt}\rightarrow c_{t}(r-r_{h})~, \quad g_{xx} \rightarrow c_{x}~, \quad g_{rr}\rightarrow c_{r}(r-r_{h})^{-1}~.
\label{eq:asymptotics_horizon}
\ee
These conditions fix the Hawking temperature and the entropy density of the bulk geometry:
\be
 T = \frac{1}{4\pi}\sqrt{\frac{c_{t}}{c_{r}}}~, \quad s = \frac{c_{x}^{(d-1)/2}}{4G_{d+1}}~.
 \label{eq:s-wave_t_s}
\ee
The model exhibits a second-order phase transition. At high temperatures, the scalar field $\Psi$ vanishes and one obtains the standard Reissner-Nordstr\"{o}m-AdS black hole. But at low temperatures, the Reissner-Nordstr\"{o}m-AdS black hole becomes unstable and is replaced by a charged black hole with a scalar ``hair." 

This system is supposed to be dual to some kind of superconductors/superfluids. In fact, the low temperature phase shows the expected behavior for superconductors/superfluids. As superconductors, one can see the divergence of the DC conductivity, an energy gap proportional to the size of the condensate, and the holographic London equation \cite{Hartnoll:2008vx,Hartnoll:2008kx,Maeda:2008ir,Maeda:2010br}. As superfluids, one can see the existence of the second and fourth sounds \cite{Herzog:2009ci,Yarom:2009uq}. Also, vortex solutions have been constructed \cite{Albash:2009ix,Albash:2009iq,Montull:2009fe,Maeda:2009vf,Keranen:2009re}.

\subsection{$\eta/s$}

Since we are interested in the viscosity, the main object to study is the boundary energy-momentum tensor. According to the standard AdS/CFT dictionary \cite{Maldacena:1997re,Witten:1998qj,Witten:1998zw,Gubser:1998bc}, the bulk gravitational perturbations act as the source for the boundary energy-momentum tensor. Thus, our task amounts to solve the bulk gravitational perturbations. 

Consider the fluctuations of the energy-momentum tensor $T_{\mu\nu}$ which behaves as $e^{-i\omega t}$. The fluctuations are decomposed by the little group $SO(d-1)$ acting on $x_i$ ($i=1, \cdots, d-1$). The fluctuations are decomposed as the tensor mode, the vector mode (``shear mode"), and the scalar mode (``sound mode"). 

One can use various methods to compute the shear viscosity. Among them, the most powerful one is the Kubo formula method, which uses the tensor mode:
\be
\eta = - \lim_{\omega\rightarrow 0} \frac{1}{\omega} {\rm Im}\, G_R^{1212}(\omega, \vec{0})~,
\label{eq:kubo}
\ee
where $G_R^{1212}(\omega, \vec{0})$ is the retarded Green function for the tensor mode $T^{12}$ at zero spatial momentum:
\be
G_R^{1212}(\omega, \vec{0}) = -i \int^{\infty}_{-\infty} d^dx\, e^{i\omega t} 
\theta(t)  \left\bra [T^{12}(t,\vecx), T^{12}(0,\vec{0}) ] \right\ket~.
\label{eq:response_emtensor}
\ee

To obtain the retarded Green function, we consider homogeneous gravitational perturbations which take the form
\be
g_{MN}=g^{(0)}_{MN} + h_{MN}~,
\ee
where $g^{(0)}_{MN}$ is the background metric (\ref{eq:s-wave_metric}). In the Lorentzian prescription of the AdS/CFT duality \cite{Son:2002sd}, the retarded Green function (\ref{eq:response_emtensor}) can be calculated from the tensor mode $h_{12}$. 
We expand the action in terms of $\phi(t,r) := h^{1}{}_{2}(t,r)$ up to quadratic order and use the Fourier transformation
\be
 \phi(t,r) = \int \frac{d^dk}{(2\pi)^d} e^{-i\omega t + i\vec{k}\cdot \vec{x}}f_{k}(r)\tilde{\phi}_{0}(k). 
 \label{eq:Fourier}
\ee
The retarded Green function is obtained as follows: 
\begin{enumerate}

\item Solve the classical equation of motion for the field $f_{k}(r)$ with the ingoing-wave boundary condition at the horizon and $f_{k}(r) \rightarrow 1$ at the boundary. \\

\item Substitute the classical solution into the action and represent the action in terms of the boundary value $\tilde{\phi}_{0}$. Only surface terms remain, and drop the contribution from the horizon.

\item The retarded Green's function is given by the kernel of the on-shell action:
\be
 S_{\text{on-shell}} = -\frac{1}{2}\int \frac{d^{d}k}{(2\pi)^{d}} \tilde{\phi}_{0}(-k) G_{R}(k)\tilde{\phi}_{0}(k)
\label{eq:GR}
\ee
where the on-shell action is defined as $S_{\text{on-shell}} = (S + S_{\text{GH}} + S_{\text{c.t.}})|_{\text{on-shell}}$. $S_{\text{GH}}$ is the Gibbons-Hawking term to provide a correct variational problem for the background geometry. $S_{\text{c.t.}}$ is the counterterm to renormalize divergences in the classical action.
\end{enumerate}
Thus, the problem is to solve the equation of motion for the field $\phi$ under the appropriate boundary conditions.

From \eq{action_s-wave}, the action which is quadratic in $\phi$ is
\be
\begin{split}
  {}^{(2)}\! S_{s}= \frac{1}{16\pi G_{d+1}}\int d^{d+1}x
	\Bigg[
		& -\frac{1}{2}\srmgz (\nabla_{M}\phi)^{2}  \\
		& +\del_{r}\left\{
			\srmgz\left(2 g^{rr}\phi\del_{r}\phi + 
			\frac{1}{2}\frac{g_{xx}'}{g_{xx}}g^{rr}\phi^{2}\right)
		\right\}
	\Bigg]~,
\end{split}
\label{eq:action_quadratic}
\ee
with the help of background equation of motions (See \appen{s-wave_quad}). Because of the little group $SO(d-1)$ acting on $x_{i}$, the tensor mode of the metric perturbations decouples from the rest of perturbations: the other modes of the metric perturbations $h_{MN}$, the gauge field perturbations $\delta A_{M}$ and the scalar field perturbation $\delta\Psi$. Thus, they can be set to zero consistently. Since the background geometry must satisfy the stationary condition, we add the Gibbons-Hawking term 
\be
 S_{\text{GH}} = \frac{1}{16\pi G_{d+1}}\int _{r \rightarrow \infty} d^{d}x\sqrt{-\gamma}2K~,
\ee
where $\gamma^{\mu\nu}$ is the boundary induced metric, $n_{M}$ is the normal vector to the boundary and $K=\gamma^{\mu\nu}\nabla_{\mu}n_{\nu}$ is the trace of the extrinsic curvature of the boundary. 
This provides surface terms
\be
 {}^{(2)}\! S_{\text{GH}} = \frac{1}{16\pi G_{d+1}}\int_{r\rightarrow\infty} d^{d}x 
	\left(
	-2\srmgz g^{rr} \phi\del_{r}\phi - \frac{1}{\sqrt{g_{rr}}}\del_{r}\left(\sqrt{-\gamma^{(0)}}\right)\phi^{2}
	\right) ~.
\ee
Therefore, the bare action is
\be
\begin{split}
 {}^{(2)}\! \left(S_{s}+S_{\text{GH}}\right) = &\frac{1}{16\pi G_{d+1}}\int d^{d+1}x \srmgz \left[-\frac{1}{2}\left(\nabla_{M}\phi\right)^{2}\right]\\
 &+\frac{1}{16\pi G_{d+1}}\int_{r\rightarrow\infty}d^{d}x 
	\left(
		\frac{g'_{xx}}{2g_{xx}}\frac{\sqrt{-\gamma^{(0)}}}{\sqrt{g_{rr}}}
		-\frac{1}{\sqrt{g_{rr}}}\del_{r}\left(\sqrt{-\gamma^{(0)}}\right)
	\right)\phi^{2}~.
\end{split}
\label{eq:s-wave_bareaction}
\ee
The action diverges as $r \rightarrow \infty$, so the counterterms at the boundary must be added. We need only the gravitational counterterm in order to evaluate the retarded Green's function for the energy-momentum tensor. According to the holographic renormalization procedure, the counterterm is
\be
\begin{split}
 S_{\text{c.t.}} = -\frac{1}{16\pi G_{d+1}}&\int_{r\rightarrow\infty}d^{d}x \sqrt{-\gamma}
	\bigg[
		\frac{2(d-1)}{L} 
		+ \frac{L}{d-2}\mathcal{R}[\gamma]\\
		&+\frac{L^3}{(d-4)(d-2)^{2}}\left(\mathcal{R}_{\mu\nu}[\gamma]\mathcal{R}^{\mu\nu}[\gamma]-\frac{d}{4(d-1)}\mathcal{R}[\gamma]^{2}\right)
		+ \cdots
	\bigg]~,
\end{split}
\label{eq:counter_term}
\ee
where $L$ is the AdS radius and $\mathcal{R}_{\mu\nu}[\gamma]$ is the Ricci tensor made from the induced metric $\gamma_{\mu\nu}$. These terms largely depend on the spacetime dimensions%
\footnote{One has to be careful when the number of the boundary spacetime dimensions $d$ is an even number. 
See Ref.~\citen{deHaro:2000xn} for details.}. 
However, in order to evaluate the shear viscosity, we need boundary terms only up to first order in $\omega$: $\cO(\omega^2)$ terms in the action do not contribute to the Kubo formula because of the $\omega\rightarrow 0$ limit. So, only the first term in \eq{counter_term} is important and it becomes
\be
{}^{(2)}\! S_{\text{c.t.}} = \frac{1}{16\pi G_{d+1}}\int_{r\rightarrow \infty}d^{d}x \sqrt{-\gamma^{(0)}}(d-1)\phi^{2}~, 
\label{eq:counter_term_quad}
\ee
for the tensor mode perturbation. This term removes the divergences from the second term of \eq{s-wave_bareaction}. As a result, the renormalized action is
\be
\begin{split}
 16\pi G_{d+1}{}^{(2)}\! &\left(S_{s} + S_{\text{GH}} + S_{\text{c.t.}}\right)  \\
   =&\int \frac{d^dk}{(2\pi)^d} \tilde{\phi}_{0}(-k)\left(-\frac{1}{2} \frac{\srmgz}{g_{rr}}f_{-k}(r)\del_r f_{k}(r) \right)\tilde{\phi}_{0}(k)\Bigg{|}_{r\rightarrow \infty} \\
  &+ (\text{terms which are proportional to the EOM}) \\ &+ (\text{contact terms})~,
\end{split}
\label{eq:s-wave_total_action}
\ee
Here, we neglected the second derivative respect to t because it provide only $\cO\left(\omega^2\right)$ terms. ``(contact terms)" provide contact terms in the Green function and have the form $f_{-k}f_{k}$. 
They will not affect the shear viscosity since they do not give an imaginary part of retarded Green's function. The terms which give the imaginary part take the form like $f_{-k}\del_{r}f_{k}$
\footnote{So, the second term of \eq{s-wave_bareaction} and the counterterm (\ref{eq:counter_term_quad}) do not affect the shear viscosity.}. 
We will see this at the end of this section. With (\ref{eq:Fourier}) 

In order to find the on-shell action, we need to solve the equation of motion for $f_{k}(r)$:
\be
 f_{k}'' + \frac{g_{rr}}{g_{tt}}\omega^{2}f_{k} + \frac{\left(g^{rr}\srmgz\right)'}{g^{rr}\srmgz}f_{k}' =0~,
\label{eq:eom_for_tensormode}
\ee
where the long wavelength limit $\vec{k}\rightarrow 0$ is taken since $\cO(|\vec{k}|)$ terms in the action don't contribute to the Kubo formula. The equations of motion can be solved as follows.
First, solve this equation of motion near the horizon and impose the ingoing-wave boundary condition. Second, find the solution over the whole region in the bulk up to first order in $\nw$. Finally, match these solutions.


First, consider the near-horizon limit of \eq{eom_for_tensormode}. 
With asymptotics of the metric (\ref{eq:asymptotics_horizon})
\be
 f_{k}(r) \sim \left(r/r_{h}-1\right)^{\pm i\nw} = \exp\Big[\pm i \nw\ln\left[r/r_{H}-1\right]\Big]~.
\ee
where $\nw := \omega/4\pi T$ is the rescaled dimensionless frequency. 
The ingoing-wave solution is given by $f_{k}(r) = \exp\left[-i \nw \ln\left(r/r_{h}-1\right)\right]$. We expand this solution in terms of $\nw \ln\left(r/r_{h}-1\right)$ near the horizon since we take the $\nw\rightarrow0$ limit at the end of the analysis. So, 
\be
 f_{k}(r) \sim 1-i\nw \ln \left[r/r_{h}-1\right]~.
 \label{eq:near_horizon_solution}
\ee
is the boundary condition as $r \rightarrow r_{h}$. The overall factor will be determined by the boundary condition at $r\rightarrow\infty$.

Next, get the solution of \eq{eom_for_tensormode} for all $r$. In order to evaluate the Kubo formula, it is enough to obtain $f_{k}(r)$ up to first order in $\nw$. Thus, expand $f_{k}(r)$ in power of $\nw$:
\be
 f_{k}(r) = f^{(0)}(r)+\nw f^{(1)}(r) + \cO(\nw^{2})~.
\ee
Inserting this into the equation of motion, $f^{(0)}$ and $f^{(1)}$ satisfy
\be
 f^{(i)}{}'' 
 + \frac{\left(g^{rr}\srmgz\right)'}{g^{rr}\srmgz}f^{(i)}{}' = 0~, 
\ee
where $i$ runs $i=0,1$. Solutions are given by
\be
 f^{(i)}(r) = C^{(i)}_{1}+C^{(i)}_{2}\int^{\infty}_{r}dr'
  \frac{g_{rr}(r')}{\sqrt{-g^{(0)}(r')}}~,
 \label{eq:tensormode_solutions}
\ee
where $C^{(i)}_{j}$'s are integration constants. From the boundary condition at $r \rightarrow \infty$, 
\be
 C^{(0)}_{1}=1~,\quad C^{(1)}_{1} = 0~.
\ee
The rest of constants are determined by the boundary condition at the horizon. Since the integrand in \eq{tensormode_solutions} has a simple pole at the horizon, 
\be
 \int^{\infty}_{r\sim r_{h}} dr' \frac{g_{rr}(r')}{\sqrt{-g^{(0)}(r')}} 
 \sim \sqrt{\frac{c_{r}}{c_{t}c_{x}^{d-1}}} \int^{\infty}_{r/r_h\sim1} \frac{d(r'/r_{h})}{(r'/r_{h}-1)} 
 = -\frac{1}{16\pi G_{d+1} sT} \ln\left[r/r_{h}-1\right]~.
\ee
Comparing this with the boundary condition (\ref{eq:near_horizon_solution}), one gets
\be
 C^{(0)}_{2}=0, \quad \nw C^{(1)}_{2} =  4G_{d+1} \cdot i\omega s~.
\ee
Therefore, the solution of \eq{eom_for_tensormode} with the appropriate boundary conditions is
\be
 f_{k}(r) = 
 \left(1+ 4G_{d+1} \cdot i\omega s \int^{\infty}_{r}dr'\frac{g_{rr}(r')}{\sqrt{-g^{(0)}(r')}}
 + \cO \left( \nw^{2} \right)\right)~,
\label{eq:s-wave_sol}
\ee
which becomes
\be
f_{k}(r) \rightarrow 1 , \quad 
\del_{r} f_{k}(r) \rightarrow - 4G_{d+1} \cdot i\omega s \frac{g_{rr}(r)}{\sqrt{-g^{(0)}(r)}}~,
\label{eq:asymptotics}
\ee
as $r\rightarrow \infty$. Thus, the terms $f_{-k}f_{k}$ and $f_{-k}\del_{r}f_{k}$ in the action provide real and imaginary parts, respectively. So, the contact terms, which have the form $f_{-k}f_{k}$, do not contribute to the shear viscosity.

Now, we are ready to extract the Green function from the on-shell action. Substituting the solution (\ref{eq:s-wave_sol}) into the on-shell action, one gets
\be
\begin{split}
 S_{\text{on-shell}} 
 &=\frac{1}{16\pi G_{d+1}}
  \int \frac{d^dk}{(2\pi)^d} \tilde{\phi}_{0}(-k)\left(-\frac{1}{2} \frac{\srmgz}{g_{rr}}f_{-k}(r)\del_r f_{k}(r) \right)\tilde{\phi}_{0}(k)\Bigg{|}_{r\rightarrow \infty}\\
 &=-\frac{1}{2}\int \frac{d^{d}k}{(2\pi)^d} \tilde{\phi}_{0}(-k) \left(-\frac{i\omega s}{4\pi}\right)\tilde{\phi}_{0}(k)
\end{split}
\ee
This leads to the retarded Green's function
\be
G_{R}^{1212}(\omega,0) 
=-\frac{i\omega s}{4\pi} + \cO\left(\left(\omega/T\right)^{2}\right)~,
\label{eq:sGR}
\ee
from the prescription (\ref{eq:GR}). Finally, the Kubo formula (\ref{eq:kubo}) derives the shear viscosity, 
\be
 \eta = -\lim_{\omega\rightarrow 0}\frac{1}{\omega}G_{R}^{1212}(\omega,\vec{0})= \frac{s}{4\pi}~.
\ee
Thus, the ratio of the shear viscosity to the entropy density is
\be
 \frac{\eta}{s} = \frac{1}{4\pi}~.
\ee
Therefore, the universality of $\eta/s$ holds in this system irrespective of whether the complex scalar condenses or not.

\section{Anisotropic superfluids}\label{sec:p-wave}

The $p$-wave or the $(p+ip)$-wave holographic superfluids are described by the Einstein-Yang-Mills system:
\be
S_{\text{EYM}} = \frac{1}{16\pi G_{d+1}}\int d^{d+1}x \sqrt{-g} \left\{ 
       R - 2 \Lambda - \frac{1}{4} (F_{MN}^a)^2 \right\}~,
\label{eq:action_p-wave}
\ee
where 
$F_{MN}^a = \del_M A_N^a - \del_N A_M^a + g_{\text{YM}} \epsilon^{abc} A_M^b A_N^c$
is the field strength of $SU(2)$ gauge fields, $g_{\text{YM}}$ is the Yang-Mills gauge coupling and $\epsilon^{abc}$ is the totally antisymmetric tensor with $\epsilon^{123}=1$. The gauge field is written as a matrix-valued form:
\be
A = A_M^a \tau^a dx^M~,
\ee
where $\tau^a = \sigma^a/(2i)$ using the Pauli matrices, so 
$[\tau^a, \tau^b]= \epsilon^{abc} \tau^c$.

\subsection{The $p$-wave superfluids}

The $p$-wave case is described by the ansatz
\bea
ds^2_{d+1} &=& - g_{tt}(r) dt^2 
+ g_{x_1 x_1}(r) dx_1^2+ g_{x_2 x_2}(r) \sum_{i=2}^{d-1} dx_i^2 + g_{rr}(r) dr^2~, 
\label{eq:metric_p}\\
A &=& \Phi(r) \tau^3 dt + w(r) \tau^1 dx_1~.
\eea
The function $\Phi(r)$ gives the background static electric potential whereas the function $w(r)$ represents the condensate. We impose the regularity condition at the horizon $r=r_{h}$:
\be
 g_{tt} \rightarrow c_{t}(r-r_{h})~, \quad g_{x_{1}x_{1}} \rightarrow c_{x_{1}}~, \quad g_{x_{2}x_{2}} \rightarrow c_{x_{2}}~, \quad 
 g_{rr} \rightarrow c_{r}(r-r_{h})^{-1}~.
\ee
Then, the temperature and the entropy density are given by
\be
 T = \frac{1}{4\pi} \sqrt{\frac{c_{t}}{c_{r}}}~, \quad s = \frac{\sqrt{c_{x_{1}}c_{x_{2}}^{d-2}}}{4G_{d+1}}~,
\ee
respectively.

As is clear from the metric (\ref{eq:metric_p}), the boundary spacetime is anisotropic. In such a case, the shear viscosity is no longer given by a single coefficient $\eta$. Rather, one is interested in
\bea
\eta^{ijkl} &=& - \lim_{\omega\rightarrow 0} \frac{1}{\omega} {\rm Im}\, G_R^{ijkl}(\omega, \vec{0})~,
\\
G_R^{ijkl}(\omega, \vec{0}) &=& -i \int^{\infty}_{-\infty} d^dx\, e^{i\omega t} 
\theta(t)  \left\bra [T^{ij}(t,\vecx), T^{kl}(0,\vec{0}) ] \right\ket~.
\eea
From the symmetric nature of the metric and the $SO(d-2)$ invariance acting on $x_2, \cdots, x_{d-1}$, there are only two nontrivial independent coefficients, {\it e.g.}, $\eta^{1212}$ and $\eta^{2323}$. We will examine these coefficients below.

The shear viscosities of anisotropic fluids have been widely discussed in the context of liquid crystal \cite{liquid_crystal}. There are various parametrizations known in the literature. Among them, the most well-studied parametrization is the Miesowicz viscosity coefficients. The coefficients $\eta^{1212}$ and $\eta^{2323}$ are related to the Miesowicz coefficients \cite{sarman_evans}. However, various conventions are found in the literature for the Miesowicz coefficients. To avoid the confusion, we keep using $\eta^{ijkl}$. 

\subsubsection{$\eta^{2323}$ (for $d \geq 4$)}

First, let us consider $\eta^{2323}$. The coefficient exists for $d \geq 4$, and the metric has the $SO(d-2)$ invariance, so the perturbation $h_{23}$ transforms as a tensor mode. Then, the discussion is similar to the $s$-wave superfluid case.
 
The action (\ref{eq:action_p-wave}) with appropriate boundary terms reduces to ($h^{2}_{~3} =: \phi(r,t) = \int \frac{d^{d}k}{(2\pi)^d} e^{ikx}f_{k}(r)\tilde{\phi}_{0}(k)$)
\be
\begin{split}
 16\pi G_{d+1}{}^{(2)}\! &\left(S_{p} + S_{\text{GH}} + S_{\text{c.t.}}\right)  \\
   =&\int \frac{d^dk}{(2\pi)^d} \tilde{\phi}_{0}(-k)\left(-\frac{1}{2} \frac{\srmgz}{g_{rr}}f_{-k}(r)\del_r f_{k}(r) \right)\tilde{\phi}_{0}(k)\Bigg{|}_{r\rightarrow \infty} \\
  &+ (\text{terms which are proportional to the EOM}) \\ &+ (\text{contact terms})~,
\end{split}
\ee
using the ansatz (\ref{eq:metric_p}) and the equation of motion for the background geometry (See \appen{p-wave_quad}). The equation of motion is given by
\be
 f_{k}'' + \frac{g_{rr}}{g_{tt}}\omega^{2}f_{k} +\frac{\left(g^{rr}\srmgz\right)'}{g^{rr}\srmgz} f_{k}' =0~.
\label{eq:p-eom_for_tensormode}
\ee
This takes the same form as the $s$-wave case (\ref{eq:eom_for_tensormode}), so the solution under the appropriate boundary conditions is given by
\be
 f_{k}(r) = \left(1+ 4G_{d+1} \cdot i \omega s \int^{\infty}_{r}dr' \frac{g_{rr}(r')}{\sqrt{-g^{(0)}(r')}} 
 + \cO \left( \left( \omega/T \right)^{2} \right)\right)~,
\ee
and the retarded Green function has the same form as the $s$-wave one (\ref{eq:sGR}). From the Kubo formula, the shear viscosity to the entropy density ratio is
\be
 \frac{\eta^{2323}}{s} = \frac{1}{4\pi}~.
\ee
The universality holds for this case as well.

\subsubsection{$\eta^{1212}$}

Next, let us consider $\eta^{1212}$. For $d = 3$, this is the only shear viscosity coefficient. The perturbation $h_{12}$ transforms as a vector under $SO(d-2)$. Thus, it couples to the vector mode of the Yang-Mills perturbations. As a result, the existing technique is not applicable. 

It should be straightforward to obtain the action for the relevant vector mode perturbations since they are standard vector perturbations for which one can rely on the symmetry. However, it does not seem straightforward to solve them analytically. Thus, we will not discuss those perturbations further in this case. This is in contrast to the $(p+ip)$-wave case in next subsection, where it is not very clear how the relevant modes are coupled. This is because the symmetry structure is more complicated for the $(p+ip)$-wave case.

\subsection{The $(p+ip)$-wave superfluids}

For completeness, let us consider the $(p+ip)$-wave holographic superfluid. The $(p+ip)$-wave case is described by the ansatz
\bea
ds^2 &=& - g_{tt}(r) dt^2 + g_{xx}(r) (dx_1^2 + dx_2^2) + g_{rr}(r) dr^2~, 
\label{eq:metric_p+ip} \\
A &=& \Phi(r) \tau^3 dt + w(r) (\tau^1 dx_1 + \tau^2 dx_2)~.
\label{eq:gauge_p+ip}
\eea
As in previous models, the regularity condition at the horizon $r=r_{h}$ implies
\be
 g_{tt} \rightarrow c_{t}(r-r_{h})~, \quad g_{xx} \rightarrow c_{x}~, \quad g_{tt} \rightarrow c_{t}(r-r_{h})^{-1}~,
\ee
which leads to the temperature and the entropy density as
\be
 T = \frac{1}{4\pi}\sqrt{\frac{c_{t}}{c_{r}}}~, \quad s = \frac{c_{x}}{4 G_{4}}~,
\ee
respectively.
It is argued that the $(p+ip)$-wave background is unstable and it turns into the $p$-wave background \cite{Gubser:2008wv}. But the analysis was carried out only in the probe limit and the full analysis including the backreaction has not been done.

Unlike the $p$-wave case, the metric is isotropic. Thus, the shear viscosity is described by a single coefficient. The anisotropy in the $(x_1,x_2)$-plane is caused by the Yang-Mills field. The condensation breaks the $SO(2)$ rotational symmetry in the $(x_1,x_2)$-plane as well as the $U(1)$ gauge symmetry. But, as is clear from \eq{gauge_p+ip},  it preserves a diagonal $U(1)$ which is a combination of the two. Thus, there does not exist the tensor mode which decouples from Yang-Mills perturbations. As a result, the existing technique is not applicable. 

Since the whole system does not have the $SO(2)$ symmetry, it is worthwhile to see explicitly how Yang-Mills perturbations couple to the ``tensor mode" perturbations. In \appen{EYM_quad}, we derived the interaction of the Yang-Mills perturbations and the metric perturbations [\eq{EYM_quad_int}]. For the tensor mode metric perturbations $h^{2}{}_{2}=-h^{1}{}_{1} (:=\phi_{d})$ and $h^{1}{}_{2}=h^{2}{}_{1} (:=\phi_{od})$%
\footnote{In the $s$-wave and $p$-wave cases, the diagonal perturbation $\phi_{od}$ and the off-diagonal perturbation $\phi_{d}$ are completely decoupled. So, we have set $\phi_{d}=0$. But this does not hold for the $(p+ip)$-wave case as we will see below.}, 
the interaction reduces to
\be
 \frac{1}{2}h_{ij}Q^{ijMN}_{a}f^{a}_{MN} 
 = h^{i}{}_{j}\left(F\cdot f\right)^{j}{}_{i}~.
 \label{eq:EYM_quad_int_text}
\ee
Here, we defined 
\bea
 f^{a}_{MN} &:=& D_{M}a^{a}_{N}-D_{N}a^{a}_{M}~,\\
 D^{ab}_{M} &:=& \nabla_{M} \delta^{ab} + g_{\text{YM}}\epsilon^{acb}A^{c}_{M}~,\\
%
%
 (F\cdot f)^{i}{}_{j} &:=& F^{aiN}f^{a}_{jN}~.
\eea

As is clear from \eq{EYM_quad_int_text}, the tensor mode $h^{i}{}_{j}$ couples to $(F\cdot f)^{i}{}_{j}$, which transforms as a tensor under the diagonal $U(1)$ symmetry. $\phi_{d}$ and $\phi_{od}$ couple to
\be
(F\cdot f)_{d} := (F\cdot f)^{2}{}_{2}=-(F\cdot f)^{1}{}_{1}~,\qquad 
(F\cdot f)_{od}:= (F\cdot f)^{1}{}_{2}= (F\cdot f)^{2}{}_{1}~,
\ee
respectively.
They contain 
the following components of $\delta A^a_{~M}$:
\be
 a_{d} := \delta A^{1}{}_{1} = -\delta A^{2}{}_{2}~,\qquad
 a_{od} := \delta A^{1}{}_{2} = \delta A^{2}{}_{1}~,\qquad
 (\text{the other modes}) = 0~.
\label{eq:adaod}
\ee

These perturbations $\phi_{od}$, $\phi_{d}$,  $a_{od}$ and $a_{d}$ are all coupled. The explicit form of $(F\cdot f)$ is given by
\bea
 (F\cdot f)_{od} &=& g^{rr}g^{xx} (\del_{r}w)(\del_{r}a_{od}) + g^{tt}g^{xx}g_{\text{YM}}\Phi w (D_{t}a_{d}), \\
 (F\cdot f)_{d}  &=& g^{rr}g^{xx} (\del_{r}w)(\del_{r}a_{d})  - g^{tt}g^{xx}g_{\text{YM}}\Phi w (D_{t}a_{od}),
\eea
which include the covariant derivatives of $a_{i}$:
\be
 D_{t}a_{od} = \del_{t}a_{od} + g_{\text{YM}} \Phi a_{d}~, \qquad D_{t}a_{d} = \del_{t}a_{d} - g_{\text{YM}} \Phi a_{od}~.
\ee
Therefore, these forms mix $a_{od}$ and $a_{d}$.

Let us summarize how the tensor mode metric perturbations couple with Yang-Mills perturbations:
\begin{itemize}
 
 \item The tensor mode metric perturbations $\phi_{od}$ and $\phi_{d}$ couple to the tensor $(F\cdot f)_{od}$ and $(F\cdot f)_{d}$, respectively, where $(F\cdot f)$ is made from the Yang-Mills perturbations.
So, the action for $\phi_{od/d}$ no longer takes the minimally-coupled scalar form. 
 
 \item The Yang-Mills perturbations $a_{od/d}$ couple to $\phi_{od}$ and $\phi_{d}$. As a result, $\phi_{od}$ couples to $\phi_{d}$ through $a_{i}$. 
 
\end{itemize}

The complete action in terms of these ``tensor mode" fluctuations are given by
\bea
 {}^{(2)}\! (S_{(p+ip)}&+&S_{\text{GH}}) = 
 \frac{1}{16\pi G_{4}}\int d^{4}x 
 {}^{(2)}\!\left(\mathcal{L}_{\text{grav}} + \mathcal{L}_{\text{gauge}} + \mathcal{L}_{\text{int}}\right)~; \\
 {}^{(2)}\!\mathcal{L}_{\text{grav}} &=& \srmgz \sum_{i=1}^{2}
	\bigg[
	-\frac{1}{2}\left\{
		-g^{tt}(\del_{t}\phi_{i})^{2}
		+g^{rr}(\del_{r}\phi_{i})^{2}
	\right\}
 -\frac{1}{2}M(r)^{2}\phi_{i}^{2} \notag \\ & & \qquad\qquad\qquad  +(\text{surface term})
	\bigg]~, \\
 {}^{(2)}\!\mathcal{L}_{\text{gauge}} &=& 
   \srmgz g^{xx} \sum^{2}_{i=1} 
	\left[
		 -g^{rr}(\del_{r}a_{i})^{2} + g^{2}_{\text{YM}}w^{2}a^{2}_{i} + g^{tt}(D_{t}a_{i})^{2} 
	\right]~, \\
 {}^{(2)}\!\mathcal{L}_{\text{int}} &=& 
  \srmgz\sum^{2}_{i=1} \phi_{i}(F\cdot f)_{i}~;
\eea
where we defined two-component vectors as $\phi_{i}= (\phi_{od}, \phi_{d})$, $a_{i}= (a_{od}, a_{d})$, $(F\cdot f)_{i}=\left((F\cdot f)_{1}, (F\cdot f)_{2}\right)$ and $i$ runs isotropic components $i=1,2$.
$ {}^{(2)}\!S_{\text{int}}$ is the interaction term we have discussed in \eq{EYM_quad_int_text}. 
The mass-like function $M(r)$ is defined by 
\be
 M(r)^{2} := g^{rr}g^{xx} (\del_{r}w)^{2} + g^{xx}g^{2}_{\text{YM}}\left(g^{xx}w^{2}-g^{tt}\Phi^{2}\right)w^2~.
\ee

The action leads to coupled equations of motion for $\phi_{od}$, $\phi_{d}$,  $a_{od}$ and $a_{d}$. It is difficult to solve them analytically, and it does not seem straightforward to obtain the shear viscosity to the entropy ratio.

\section{Implications of the results}\label{sec:discussion}



We study $\eta/s$ for $s$-wave, $(p+ip)$-wave, and $p$-wave holographic superfluids. The shear viscosity for the $s$-wave superfluids satisfies the universality  (\ref{eq:univ}). The $p$-wave superfluids are anisotropic, and there are two nontrivial independent shear viscosities, $\eta^{2323}$ and $\eta^{1212}$. We show that one of the coefficients $\eta^{2323}$ satisfies the universality.

On the other hand, for another coefficient $\eta^{1212}$ of the $p$-wave superfluids and for the shear viscosity of the $(p+ip)$-wave superfluid, the gravitational perturbations in question couple to the Yang-Mills perturbations even in Kubo-formula method, and the existing technique is not applicable. We extract the modes which couple to the gravitational perturbations. For the $(p+ip)$-wave case, we write down the perturbed action for those modes. 

It would be interesting to solve the equations of motion to obtain $\eta/s$ for those cases.
If it turns out that they also satisfy the universality, these shear viscosities will give highly nontrivial tests for the universality. If it turns out that they do not, these shear viscosities will give interesting counterexamples against the universality. The results are interesting in either way.

There is another technique to derive $\eta/s$ in the membrane paradigm context \cite{Iqbal:2008by}. According to the method, transport coefficients in the boundary field theory can be determined by (i) the flow equation for $r$-dependent transport coefficients, {\it e.g.}, the shear viscosity $\eta(r)$ and by (ii) their values at the horizon. If the tensor mode metric perturbation is written as a free scalar, the flow equation becomes trivial in the hydrodynamic limit: $\del_{r}\eta(r) =0$.
In this case, $(\eta/s)|_{\text{boundary}}=(\eta/s)|_{\text{horizon}}=1/4\pi$. This method works for $\eta$ in the $s$-wave case and for $\eta^{2323}$ in the $p$-wave case. But it does not work for $\eta^{1212}$ in the $p$-wave case and for $\eta$ in the $(p+ip)$-wave case. This is because the interactions of the metric and Yang-Mills perturbations provide a non-trivial flow equation $\del_{r}\eta(r) \neq 0$.

We now discuss the shear viscosity itself of holographic superfluids below.


\subsection{Viscosity of superfluids}

The holographic superfluid shows a nonzero viscosity. To interpret the result, note the following points.

First, a superfluid has a nonzero viscosity. For example, for $^4$He no viscous resistance is observed when it goes through a narrow pipe, but a viscous drag is observed when a test body is moved in the liquid. 

According to the two-fluid model, a superfluid consists of the superfluid component and the normal component. The normal component has a nonzero viscosity, so a superfluid has a nonzero viscosity as a whole. The normal component represents the effect of thermal fluctuation, and it always exists at finite temperatures. And the quasi-particle description is valid for the normal component. Since we do not separate the normal and superfluid components, one cannot observe the zero viscosity for the superfluid component. 

Second, currently the boundary theory description is not clear for holographic superfluids, but the boundary theory presumably contains the fields which may not play an important role in the superfluid behavior. Among the other things, the boundary theory should include the $SU(N)$ non-Abelian gauge field, which is unlikely to play an important role. The computation of $\eta/s$ includes the dissipation not only from the normal component but also from these fields.

Obviously, while $\eta/s$ is the same in both phases, $\eta$ itself can have different functional forms. This requires the knowledge of $s$, and it would be interesting to compute it. In the probe limit, both phases are described by the same bulk geometry since the backreaction of matter fields onto the geometry is ignored. Thus, one needs the fully backreacted metric to find a nontrivial behavior. In particular, it would be interesting to see if $\eta$ in the superfluid state is lower than $\eta$ in the (unstable) normal state. Again one needs a fully backreacted metric, but analysis near the critical point or a numerical computation would suffice for the purpose. 

In fact, the entropy density $s$ has been obtained for a limited class of holographic superfluids. Especially, Ref.~\citen{Ammon:2009xh} obtained $s$ for the $(4+1)$-dimensional $p$-wave superfluid in the grand canonical ensemble.%
\footnote{For a $s$-wave superfluid, $s$ has been obtained in the microcanonical ensemble \cite{Maeda:2010hf}.}
They use the parameter $\alpha:=\kappa_5/\hat{g}$. In our notation, $\alpha \propto 1/g_{\rm YM}$, and $\alpha\rightarrow0$ corresponds to the probe limit. Once the backreaction is taken into account, the $p$-wave superfluid undergoes the second-order phase transition only when $\alpha<\alpha_c$, where $\alpha_c \sim 0.365$, and it undergoes the first-order phase transition when $\alpha>\alpha_c$. Namely, the phase transition becomes first-order when the backreaction becomes large. 

According to their computation, $s$ in the superfluid state is lower than $s$ in the unstable normal state below $T<T_c$ at fixed chemical potential $\mu=A^{3}_{t}$. See Fig.~3(b) of Ref.~\citen{Ammon:2009xh}. Since $\eta^{2323}$ satisfies the universality, $\eta^{2323}$ in the superfluid state is lower than the normal state one. This may be due to the zero viscosity of the superfluid component. Needless to say, this statement is only for one coefficient of shear viscosities of one $p$-wave superfluid. At this moment, it is not clear if the same holds in general. 

\subsection{Implication to dynamic critical phenomena}\label{sec:DCP}

We found that the universality of $\eta/s$ holds both for high temperature phase and for low temperature phase. In the second-order phase transition, critical phenomena occur and one has singular behaviors in physical quantities. In the dynamic case, one has singular behaviors in various transport coefficients \cite{Hohenberg:1977ym}. But our results indicate that there is no divergence in the shear viscosity. (Since the entropy density is the first derivative of the free energy, it is continuous across the phase transition. Thus, the universality of $\eta/s$ implies that the shear viscosity is also continuous across the transition.)

More precisely, in the dynamic critical phenomena, the relaxation time of the order parameter diverges, which is known as the critical slowing down. In fact, for $s$-wave holographic superfluids, the relaxation time of the order parameter diverges near the critical point \cite{Maeda:2009wv}. In general, when a system has a conserved charge, the associated transport coefficient diverges as well. For example, for $T_{\mu\nu}$, one has a (mild) singularity in $\eta$. But our results indicate that this does not happen in the holographic superfluids. The fact that singular behavior does not occur in $\eta$ has been observed in the critical phenomena of R-charged black holes \cite{Maeda:2008hn}.




\section*{Acknowledgements}
We would like to thank Akihiro Ishibashi, Kengo Maeda, Takashi Okamura and Shinya Tomizawa for useful discussions. The research of MN was supported in part by the Grant-in-Aid for Scientific Research (20540285) from the Ministry of Education, Culture, Sports, Science and Technology, Japan.

\appendix

\section{Quadratic forms of perturbations for Einstein-Matter actions}

In this Appendix, we derive the quadratic forms of the tensor mode perturbation for the $s$-wave, $p$-wave and $(p+ip)$-wave holographic superfluids. First, we derive the quadratic form of the Einstein-Hilbert action. Then, we derive quadratic forms of the matter action for these three models. 
Since we will not consider scalar perturbations, we will focus 
on the metric perturbations and the gauge field perturbations.

\subsection{The quadratic form of the Einstein-Hilbert action}

Consider the general perturbation $h_{MN}$ to the background metric $g^{(0)}_{MN}$:
\be
 g_{MN} = g^{(0)}_{MN} + h_{MN}~.
\ee
Under the perturbation,
\be
 \srmg = \sqrt{-g^{(0)}}\left[1+\frac{1}{2}h + \frac{1}{2}\left(\frac{1}{4}h^2-\frac{1}{2}h^{MN}h_{MN}\right)\right]~,
\ee
and 
\be
 \begin{split}
 {}^{(2)}\! R = &\nabla_{M}\left(
                h^{IJ}\nabla^{M}h_{IJ} + h^{MN}\nabla_{N}h 
                - h^{MN}\nabla_{I}h^{I}{}_{N} - h^{IJ}\nabla_{I}h_{J}{}^{M}\right) \\
             &-\frac{1}{4}(
             \nabla^{N}h^{IJ})\nabla_{N}h_{IJ} + \frac{1}{2}(\nabla^{N}h^{IJ})\nabla_{I}h_{JN}
             -\frac{1}{4}\nabla_{N}h\nabla^{N}h + {}^{(0)}\!R_{MI}h^{I}{}_{N}h^{MN}.\label{eq:R2}
 \end{split}
\ee
Therefore, the quadratic form of $h_{MN}$ in the Einstein-Hilbert action is
\be
 \begin{split}
   &\frac{{}^{(2)}\! (\srmg R)}{\srmgz} =
	\left(\frac{1}{8}h^{2}-\frac{1}{4}h^{MN}h_{MN}\right){}^{(0)}\!R 
	+\left(h^{MI}h_{I}{}^{N}-\frac{1}{2}hh^{MN}\right){}^{(0)}\!R_{MN} +\nabla_{M}J^{M}\\ 
	& -\frac{1}{4}(\nabla^{M}h^{IJ})(\nabla_{M}h_{IJ}) 
	+\frac{1}{2}(\nabla^{M}h^{IJ})(\nabla_{I}h_{JM})
	+\frac{1}{4}(\nabla_{I}h)(\nabla^{I}h)
	-\frac{1}{2}(\nabla_{J}h^{IJ})(\nabla_{I}h)~,
 \end{split}
\label{eq:EH_quad}
\ee
where
\be
 J^{M} =	h^{IJ}\nabla^{M}h_{IJ} + h^{MN}\nabla_{N}h  -h^{MN}\nabla_{I}h^{I}{}_{N} 
			- h^{IJ}\nabla_{I}h_{J}{}^{M} -\frac{1}{2}h\nabla^{M}h + \frac{1}{2}h\nabla_{N}h^{MN}~. 
\ee
If we restrict the perturbation $h_{MN}$ to the tensor mode perturbation $h_{12}$, this quadratic form reduces to ($\phi := h^{1}{}_{2}$)
\be
 {}^{(2)}(\sqrt{-g}R) = \srmgz \left[{}^{(2)}\! R -\frac{1}{2}{}^{(0)}\!R \phi^{2}\right]~,
\ee
where ${}^{(2)}\! R$ is the quadratic form of the Ricci scalar with the tensor mode perturbation:
\be
\begin{split}
 {}^{(2)}\!R &= 
	\srmgz 
		\left[
			-\frac{1}{2}(\nabla_{M}\phi)^{2}
			+{}^{(0)}\! R^{2}{}_{2} \phi^{2}
		\right] \\
		&+\del_{r}
			\left(
				2\frac{\sqrt{-g^{(0)}}}{g_{rr}}\phi\del_{r}\phi
				+\frac{\sqrt{-g^{(0)}}}{g_{rr}}\frac{g'_{x_{2}x_{2}}}{2g_{x_{2}x_{2}}}\phi^{2}
			\right)
		-\del_{t}
			\left(
				2\frac{\sqrt{-g^{(0)}}}{g_{tt}}\phi\del_{t}\phi
			\right)~.\label{eq:Rics_quad}
\end{split}
\ee
Here, we take the $(d+1)$-dimensional $p$-wave metric (\ref{eq:metric_p}) to preserve generality. 
Except the free scalar part, all these terms will be removed in the end. 

\subsection{The quadratic form of the $s$-wave holographic superfluid action}\label{sec:s-wave_quad}

The $s$-wave holographic superfluid is described by \eq{action_s-wave}:
\bea
 S_{s} &=& \frac{1}{16\pi G_{d+1}}\int d^{d+1}x ( \srmg R + \mathcal{L}_{\text{$s$-matter}})~; \\
 \mathcal{L}_{\text{$s$-matter}} &:=& \srmg\left[
 	 -\frac{1}{4}K_{1}\left(|\Psi|\right)F_{MN}F^{MN}
 	 - K_{2}\left(|\Psi|\right)|D_{M}\Psi|^{2}
 	 -V\left(|\Psi|\right)
 	 \right]~,
\eea
where we defined a covariant derivative as $D_{M}:= \nabla_{M}-iqA_{M}$.
Under the general gravitational perturbation
\be
 g_{MN} = g^{(0)}_{MN} + h_{MN}~,
\ee
one can easily find
\be
 {}^{(2)}\! \mathcal{L}_{\text{$s$-matter}} = 
		\frac{1}{2}\srmgz h_{MN}\left[K_{1}X^{MNIJ} + K_{2}Y^{MNIJ} + V P^{MNIJ}\right]h_{IJ}~,
\label{eq:s-wave_matter_quad}
\ee
where 
\bea
 P^{MNIJ} &:=& \frac{1}{4}\left(g^{(0)MI}g^{(0)NJ}+g^{(0)MJ}g^{(0)NI}-g^{(0)MN}g^{(0)IJ}\right)~,\\
 X^{MNIJ} &:=& \frac{1}{4}F_{AB}F^{AB}P^{MNIJ} +\frac{1}{2}F_{A}{}^{M}F^{AN}g^{(0)IJ}  \notag \\ && - F_{A}{}^{M}F^{AJ}g^{(0)NI}-\frac{1}{2}F^{MI}F^{NJ}~, \\
 Y^{MNIJ }&:=& |D_{A}\Psi|^{2}P^{MNIJ}
              + (D^{M}\Psi)(D^{N}\Psi)^{\ast} g^{(0)IJ} \notag \\ && -2 (D^{M}\Psi)(D^{J}\Psi)^{\ast} g^{(0)NI}~.
\eea
Note that some modes of the metric perturbations couple with the gauge field perturbations $\delta A_{I}$ and the complex scalar perturbation $\delta \Psi$ in general. However, we drop these perturbations since these decouple from the tensor mode metric perturbations. 

The equation of motion for the background field is
\be
 \left( \frac{1}{2} {}^{(0)}\! R g^{(0)MN}- {}^{(0)}\! R^{MN} \right) + T^{(0)MN} =0~.
 \label{eq:s-wave_einstein}
\ee
The background energy-momentum tensor $T^{(0)}_{MN}$ is defined as
\footnote{Here, we defined the symmetric symbol as $F^{(M}{}_{A}F^{N)A}=\frac{1}{2}\left(F^{M}{}_{A}F^{NA}+F^{N}{}_{A}F^{MA}\right)$.}
\be
\begin{split}
 T^{(0)MN} &= \frac{1}{\srmgz} \frac{\del \mathcal{L}_{\text{$s$-matter}}}{\del g _{MN}} \\
                  &= \frac{1}{2} K_{1} \left(-\frac{1}{4}g^{(0)MN}F_{AB}F^{AB} + F^{(M}{}_{A}F^{N)A}\right)\\
                  & \quad + K_{2}\left( -\frac{1}{2}g^{(0)MN}(D_{A}\Psi)(D^{A}\Psi)^{\ast} + (D^{(M}\Psi)(D^{N)}\Psi)^{\ast}\right) -\frac{1}{2}g^{(0)MN}V~.
\end{split}
\ee
This equation of motion leads to a relation between the Ricci scalar and the matter fields
\be
 {}^{(0)}\! R = \frac{d-3}{4(d-1)}K_{1}F_{MN}F^{MN} + K_{2}|D_{M}\Psi|^{2} + \frac{d+1}{d-1}V~,
 \label{eq:s-wave_eom_trace}
\ee
and an isotropic component leads to
\be
 R^{2}{}_{2} = \frac{1}{d-1} \left( V - \frac{1}{4}K_{1}F_{MN}F^{MN}\right)~.
 \label{eq:s-wave_eom_iso}
\ee

So far our discussion does not assume 
an explicit background nor perturbations. Now, we take the $s$-wave background ansatz (\ref{eq:s-wave_metric})-(\ref{eq:s-wave_scalar}) and the tensor mode $h^{1}{}_{2}=\phi(t,r)$. The quadratic form (\ref{eq:s-wave_matter_quad}) reduces to
\be
{}^{(2)}\! \mathcal{L}_{\text{$s$-matter}} =
\frac{1}{2}\srmgz 
	\left(
		\frac{1}{4}K_{1}F_{MN}F^{MN}
		+K_{2}|D_{M}\Psi|^{2}
		+V
		\right) \phi^{2}~,
\ee
where we set the gauge field perturbations $a_{M}$ to zero since these decouple from $\phi$. Using the trace of the equations of motion (\ref{eq:s-wave_eom_trace}) and an isotropic component of  \eq{s-wave_eom_iso}, one gets
\be
 {}^{(2)}\!\mathcal{L}_{\text{$s$-matter}} = \srmgz \left(\frac{1}{2}{}^{(0)}\! R - {}^{(0)}\!R^{2}{}_{2}\right)\phi^{2}~.
 \label{eq:s-wave_matter}
\ee
Combining the Einstein-Hilbert term (\ref{eq:EH_quad}) and the matter term (\ref{eq:s-wave_matter}), we obtain the quadratic form of the $s$-wave holographic superfluid action (\ref{eq:action_quadratic}):
\be
\begin{split}
  {}^{(2)}\!S_{s}= \frac{1}{16\pi G_{d+1}}\int d^{d+1}x
 \Bigg[&-\frac{1}{2}\sqrt{-g} (\nabla_{M}\phi)^{2} 
       -\del_{t}\left(2\srmgz g^{tt}\phi\del_{t}\phi\right)\\
       &+\del_{r}\left\{\srmgz \left(2g^{rr}\phi\del_{r}\phi + \frac{1}{2}\frac{g_{xx}'}{g_{xx}}g^{rr}\phi^{2}\right)
        \right\}
  \Bigg]~,
\end{split}
\ee
with $g_{x_{1}x_{1}}=g_{x_{2}x_{2}}=g_{xx}$. The second term (the total derivative with respect to $t$) does not affect the correlator, so we ignored the term in Eq. (\ref{eq:action_quadratic}).

\subsection{The quadratic form of the Einstein-Yang-Mills action}\label{sec:EYM_quad}

The Einstein-Yang-Mills action is
\bea
 S_{\text{EYM}} &=& \frac{1}{16\pi G_{d+1}}\int d^{d+1}x ( \srmg R + \mathcal{L}_{\text{EYM-matter}})~, \\
 \mathcal{L}_{\text{EYM-matter}} &:=& \srmg\left[ -\frac{1}{4}F^{a}_{MN}F^{a MN}-2\Lambda\right]~,
\eea
where $A^{a}_{M}$ is $SU(2)$ gauge field and the field strength is defined as
\be
F^{a}_{MN} = \del_{M}A^{a}_{N}-\del_{N}A^{a}_{M} + g_{\text{YM}}\epsilon^{abc}A^{b}_{M}A^{c}_{N}~.
\ee
Under the metric and gauge field perturbations 
\be
 g_{MN}=g^{(0)}_{MN}+h_{MN}~, \quad A^{a}_{M}=A^{(0)}{}^{a}_{M} + a^{a}_{M}~,
\label{eq:YMperturbation}
\ee
one can find%
\footnote{The quadratic form of the Einstein-Yang-Mills action was obtained in Ref.~\citen{Deser:1974xq} in order to calculate the one-loop divergence, but they omitted surface terms.}
\bea
 {}^{(2)}\! \mathcal{L}_{\text{EYM}} &=& 
 	{}^{(2)}\! \mathcal{L}_{\text{grav}} 
 	+{}^{(2)}\! \mathcal{L}_{\text{gauge}} 
 	+{}^{(2)}\! \mathcal{L}_{\text{int}}~;
\label{eq:EYM_quad_matter} \\ 
 {}^{(2)}\! \mathcal{L}_{\text{grav}} &=& {}^{(2)}(\srmg R)+\srmgz 
	\left[\frac{1}{2}h_{MN}\left(\bar{X}^{MNIJ}+2\Lambda P^{MNIJ}\right) h_{IJ}\right]~,\label{eq:EYM_quad_grav}\\
 {}^{(2)}\! \mathcal{L}_{\text{gauge}} &=& \srmgz 
	\left[-\frac{1}{4}f^{a}_{MN}f^{a MN}+\frac{1}{2}a^{a}_{M}Z^{MN}_{ab}a^{b}_{N}\right]~,\label{eq:EYM_quad_gauge}\\
 {}^{(2)}\! \mathcal{L}_{\text{int}} &=& 
 \frac{1}{2}\srmgz h_{MN}Q^{MNIJ}_{a}f^{a}_{IJ}~,\label{eq:EYM_quad_int}
\eea
where
\bea
 f^{a}_{MN}   &:=& D_{M}a^{a}_{N}-D_{N}a^{a}_{M} \\
 \bar{X}^{MNIJ} &:=& \frac{1}{4}F^{a}_{AB}F^{aAB}P^{MNIJ} +\frac{1}{2}F^{a}{}_{A}{}^{M}F^{aAN}g^{(0)IJ}  \\
 & & - F^{a}{}_{A}{}^{M}F^{aAJ}g^{(0)NI}-\frac{1}{2}F^{aMN}F^{aIJ}~, \notag\\
 Q^{MNIJ}_{a} &:=& 2g^{(0)MI}F^{a NJ}-\frac{1}{2}g^{(0)MN}F^{a IJ}~, \\
 Z^{MN}_{ab} &:=& -g_{\text{YM}} \epsilon^{abc}F^{cMN}~.
\eea

The background satisfies the Einstein equation (\ref{eq:s-wave_einstein}) with the energy-momentum tensor given by
\footnote{Here, the vertical bars indicate that we do not symmetrize over $a$: $F^{a(M}{}_{A}F^{|a|N)A}= \frac{1}{2}\left(F^{aM}{}_{A}F^{aNA}+F^{aN}{}_{A}F^{aMA}\right)$}
\be
 T^{(0)MN} = \frac{1}{2}\left(-\frac{1}{4}g^{(0)MN}F^{a}_{IJ}F^{aIJ}-F^{a(M}{}_{A}F^{|a|N)A}- g^{(0)MN}2\Lambda\right)~.
\ee
This equation of motion leads to a relation between the Ricci scalar and the matter fields
\be
 {}^{(0)}\! R = \frac{d-3}{4(d-1)}F^{a}_{MN}F^{aMN} + \frac{d+1}{d-1}2\Lambda ~,
 \label{eq:EYM_eom_trace}
\ee
and an isotropic component leads to
\be
 R^{2}{}_{2} = \frac{1}{d-1} \left( 2\Lambda - \frac{1}{4}F^{a}_{MN}F^{aMN}+\frac{d-1}{2}F^{a}_{M x_{2}}F^{a M x_{2}}\right)~.
 \label{eq:EYM_eom_iso}
\ee

\subsubsection{The $p$-wave holographic superfluid action (tensor mode)}\label{sec:p-wave_quad}

Here, we derive the effective action of the tensor mode metric perturbation for the $(d+1)$-dimensional $p$-wave system. If we set metric perturbation (\ref{eq:YMperturbation}) to the tensor mode $h^{2}{}_{3}=\phi(t,r)$, all the other perturbations are decoupled, so these perturbations can be ignored consistently. The quadratic action (\ref{eq:EYM_quad_matter}) reduces to
\be
 {}^{(2)}\!\mathcal{L}_{\text{EYM-matter}} 
 = \frac{1}{2}\sqrt{-g^{(0)}}\left[ \frac{1}{4}F^{a}_{MN}F^{aMN}+ 2\Lambda\right]\phi^{2}
 = \srmgz\left(\frac{1}{2}{}^{(0)}\!R-{}^{(0)}\!R^{2}{}_{2}\right)\phi^{2}~,
\ee
using Eqs.~(\ref{eq:EYM_eom_trace}) and (\ref{eq:EYM_eom_iso}). Then, one obtains the quadratic form of the $p$-wave action for the tensor mode metric perturbation:
\be
\begin{split}
 {}^{(2)}\!S_{p} = \frac{1}{16\pi G_{d+1}} \int d^{d+1}x \srmgz 
		\Bigg[
			&-\frac{1}{2}(\nabla_{M}\phi)^{2}
			-\del_{t}
			\left(
				2\sqrt{-g^{(0)}}g^{tt}\phi\del_{t}\phi
			\right)
		 \\
		&+\del_{r}
			\left\{ \srmgz
			\left(
				2g^{rr}\phi\del_{r}\phi
				+\frac{1}{2}\frac{g'_{x_{2}x_{2}}}{g_{x_{2}x_{2}}}g^{rr}\phi^{2}
			\right)
			\right\}
		\Bigg]~.
\end{split}
\ee

\subsubsection{The $(p+ip)$-wave holographic superfluid action}\label{sec:(p+ip)-wave_quad}

Let us derive the effective action of the ``tensor mode" metric perturbations for the $4$-dimensional $(p+ip)$-wave system. In this case, we must turn on four perturbations $\phi_{od}=h^{1}{}_{2}=h^{2}{}_{1}$, $\phi_{d}=h^{1}{}_{1}=-h^{2}{}_{2}$, $a_{od}=a^{1}{}_{2}=a^{2}{}_{1}$ and $a_{d}=a^{1}{}_{1}=-a^{2}{}_{2}$. 
The perturbed action is obtained by substituting these perturbations into \eq{EYM_quad_matter}:
\bea
 {}^{(2)}\! S_{(p+ip)} &=& \frac{1}{16\pi G_{4}}\int d^{4}x \sqrt{-g^{(0)}} 
	\left(
		{}^{(2)}\!\mathcal{L}_{\text{grav}}+{}^{(2)}\!\mathcal{L}_{\text{gauge}}+{}^{(2)}\!\mathcal{L}_{\text{int}}
	\right)~; \\ 
 {}^{(2)}\!\mathcal{L}_{\text{grav}} &=&  \srmgz\sum_{i=1}^{2}
	\left[
	-\frac{1}{2}\left\{
		-g^{tt}(\del_{t}\phi_{i})^{2}
		+g^{rr}(\del_{r}\phi_{i})^{2}
	\right\}
		-\frac{1}{2}M(r)^{2}\phi_{i}^{2}
	\right]~, \label{eq:p+ip_grav}\\
 {}^{(2)}\!\mathcal{L}_{\text{gauge}} &=& 
   \srmgz g^{xx} \sum^{2}_{i=1} 
	\left[
		 -g^{rr}(\del_{r}a_{i})^{2} + g^{2}_{\text{YM}}w^{2}a^{2}_{i} + g^{tt}(D_{t}a_{i})^{2} 
	\right]~, \\
 {}^{(2)}\!\mathcal{L}_{\text{int}} &=& 
   \srmgz \sum^{2}_{i=1} \phi_{i}(F\cdot f)_{i}~,
\eea
where we defined two-component vectors $\phi_{i}= (\phi_{od}, \phi_{d})$, $a_{i}= (a_{od}, a_{d})$, $(F\cdot f)_{i}=\left((F\cdot f)_{1}, (F\cdot f)_{2}\right)$ and $i$ runs isotropic components $i=1,2$. Here, we have omitted the surface term in \eq{p+ip_grav}. The mass-like function $M(r)$ is defined by 
\be
 M(r)^{2} := g^{rr}g^{xx} (\del_{r}w)^{2} + g^{xx}g^{2}_{\text{YM}}\left(g^{xx}w^{2}-g^{tt}\Phi^{2}\right)w^2~,
\ee
and the explicit form of $(F\cdot f)_{i}$ is
\bea
 (F\cdot f)_{od} &=& g^{rr}g^{xx} (\del_{r}w)(\del_{r}a_{od}) + g^{tt}g^{xx}g_{\text{YM}}\Phi w (D_{t}a_{d}), \\
 (F\cdot f)_{d}  &=& g^{rr}g^{xx} (\del_{r}w)(\del_{r}a_{d})  - g^{tt}g^{xx}g_{\text{YM}}\Phi w (D_{t}a_{od}).
\eea
Note that the covariant derivatives of $a_{i}$ have forms
\be
 D_{t}a_{od} = \del_{t}a_{od} + g_{\text{YM}} \Phi a_{d}~, \qquad D_{t}a_{d} = \del_{t}a_{d} - g_{\text{YM}} \Phi a_{od}~.
\ee
Therefore, these forms mix $a_{od}$ and $a_{d}$.


%

\end{document}